\documentclass[conference]{IEEEtran}

\usepackage{amsmath,amssymb}
\usepackage{graphicx}
\usepackage{booktabs}
\usepackage{array}
\usepackage{comment}
\usepackage{placeins}
\usepackage{flushend}

\graphicspath{{paper_figures_from_poster/}}

\begin{document}
\title{Disentangling Kinematic, Magnetic, and Solar Wind Contributions to CME Arrival Time Prediction Using a Large MHD Simulation Grid}

\author{\IEEEauthorblockN{Debosmita Mallick}
\IEEEauthorblockA{Computer Science Department\\
Georgia State University\\
Atlanta, GA, USA\\
Email: dmallick2@student.gsu.edu}
\and
\IEEEauthorblockN{Viacheslav M Sadykov}
\IEEEauthorblockA{Physics \& Astronomy Department\\
Georgia State University\\
Atlanta, GA, USA\\
Email: vsadykov@gsu.edu}
\and
\IEEEauthorblockN{Dustin Kempton}
\IEEEauthorblockA{Computer Science Department\\
Georgia State University\\
Atlanta, GA, USA\\
Email: dkempton1@gsu.edu}
\and
\IEEEauthorblockN{Elena Provornikova}
\IEEEauthorblockA{The Johns Hopkins University Applied Physics Laboratory\\
Laurel, MD, USA\\
Email: elena.provornikova@jhuapl.edu}
}

\maketitle

\begin{abstract}
Solar coronal mass ejections (CMEs) are large eruptions of plasma and magnetic field released from the Sun into the interplanetary space. When interacting with the Earth's magnetosphere, the CMEs can trigger geomagnetic storms, producing induced currents that sometimes cause significant damage to the power grid systems, and disrupt satellite communications. Despite extensive studies, the problem of predicting the time and properties of CMEs at the first Lagrange (L1) point (typically occurring at several tens of hours after the CME launch from the Sun) remains mostly unsolved, with high uncertainties ($\sim$13\,h). Additionally, the quantitative impact of different initial CME properties near the Sun, on the CME L1 arrival time, remains largely unknown. In this project, we aim to bridge this gap, by leveraging a large grid of physics-based MHD simulations, and machine learning (ML) modeling of CME arrival times. Specifically, we consider the extensive grid of the 23,058 CME simulations performed with GAMERA+Gibson-Low model. First, we isolate the solar wind background, then interpolate and subtract it from the original MHD CME simulation to recognize the CME arrival time (quantified as a consistent perturbation exceeding the interpolation accuracy). We then utilize a fully-connected neural network to predict CME arrivals at L1, based on their initial properties near the Sun and simulation boundary conditions, and quantify the impact of different CME parameters on the model. Our best result reproduces the simulated arrival times with a root-mean-square error (RMSE) of 6.47$\pm$0.48 hours ($R^2 = 0.88\pm{}0.02$), with prediction errors lower for faster CME events (in case of $v_{CME}\ge$900\,km/s). We find that the magnetic field properties of the CME have a stronger impact on arrival times with respect to the background solar wind; however, the latter is still useful in the absence of magnetic field information.
\end{abstract}


\IEEEpeerreviewmaketitle

\section{Introduction}

Space Weather refers to the dynamic activities on the Sun, its atmosphere, and the space surrounding the Earth, that can widely affect our technological systems and even our daily activities \cite{Guhathakurta2021JSWSC}. The primary drivers of space weather are solar flares, solar energetic particles (SEPs), and coronal mass ejections (CMEs). Among these, CMEs are of a special significance, as they involve the release of large quantities of magnetized plasma from the solar corona into outer space. When these are directed towards the Earth, and the CME's magnetic field direction is `favorable', the interaction between the CME and the Earth's magnetosphere can cause geomagnetic storms capable of disrupting satellite operations, radio communications, GPS navigation, aviation, and electrical power systems \cite{Schrijver2015AdSpR..55.2745S}. Predicting when a CME will reach Earth is therefore a problem of great importance in space weather research, and is essential for providing early warnings of potentially hazardous solar events.

Space weather forecasting relies on both observations and numerical simulations. Remote observations from coronagraphs aboard the Solar and Heliospheric Observatory (SOHO) and the Solar Terrestrial Relations Observatory (STEREO) provide parametric measurements of CME speed, angular width, and propagation direction near the Sun \cite{Brueckner1995SolarPhysics,Kaiser2008STEREO}. In-situ spacecraft positioned near the first Lagrange point (L1), such as the Advanced Composition Explorer (ACE) and the Deep Space Climate Observatory (DSCOVR), directly measure plasma and magnetic field parameters shortly before a CME reaches Earth \cite{Stone1998SSRvACE}. In addition, the background solar wind is typically modeled using input from magnetic field observations in the Sun`s photosphere (such as done with Helioseismic and Magnetic Imager onboard the Solar Dynamics Observatory, SDO/HMI \cite{Scherrer2012SDOHMI}). Altogether, this information can be used for modeling the propagation of CME events from Sun to Earth, including determining when they arrive at Earth.

\begin{figure*}[!htb]
        \centering
        \includegraphics[width=0.9\linewidth]{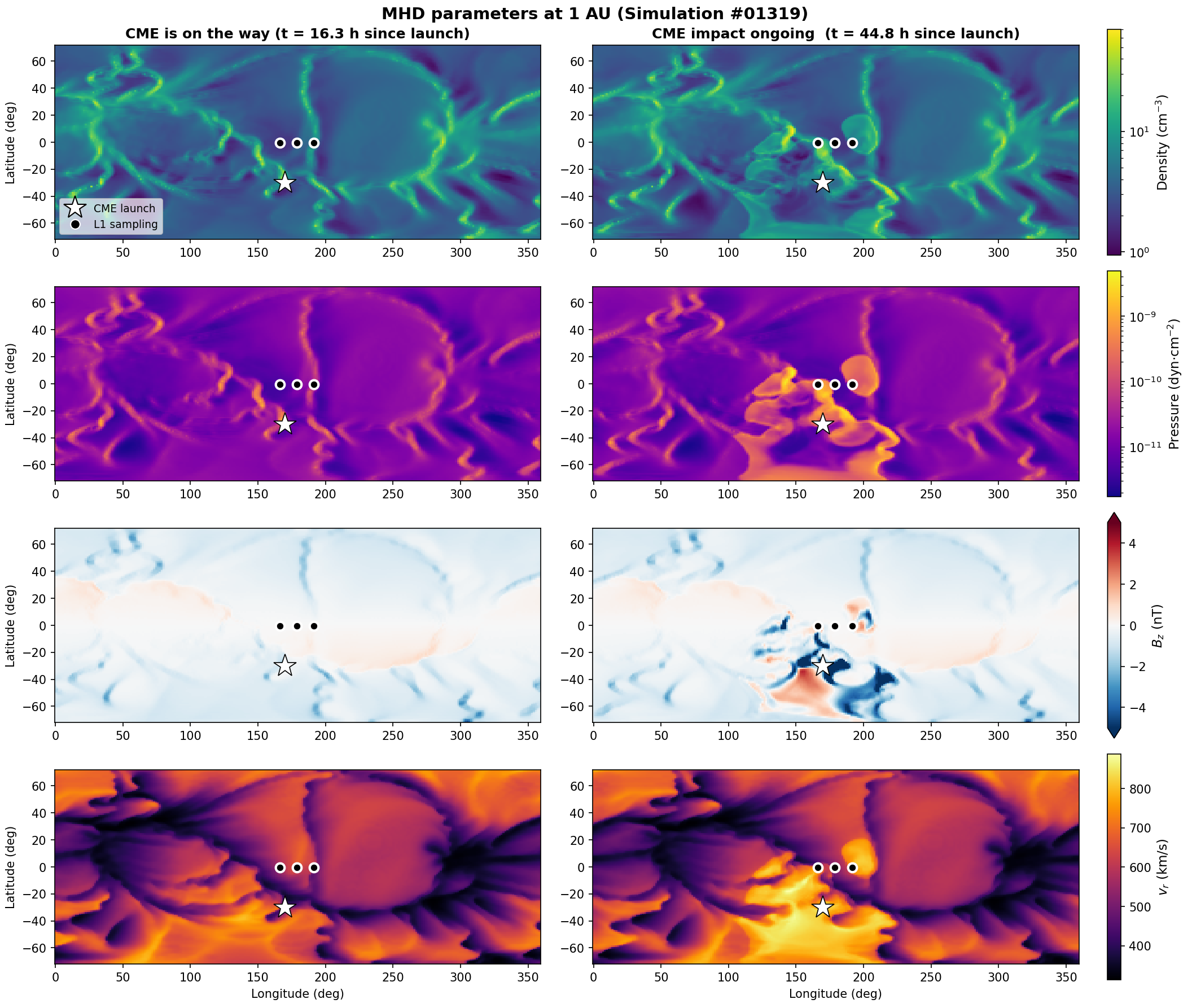}
        \caption{Snapshot of MHD parameters at 1\,AU for a representative simulated CME (run \texttt{DOICMEM\_01319}, $v_{CME}=1500$\,km/s). Left panels represent the background solar wind conditions 16.3\,h after the CME launch (and before its 1\,AU impact), right panels represent the conditions during the ongoing CME impact at 44.8\,h. The MHD properties, from top to bottom, are: plasma density, plasma pressure, magnetic field component $B_{z}$ in the GSE system, radial plasma speed $v_{r}$. The star marks the CME launch location (longitude and latitude), the black points mark the locations of sampled timeseries at 1\,AU presented in Figure~\ref{fig:bgsub}.}
        \label{fig:cme_sim}
    \end{figure*}
    
To improve prediction capabilities, researchers utilize physics-based magnetohydrodynamic (MHD) models that simulate the propagation of CMEs through the heliosphere (such as \cite{Odstrcil2003solarwind,Pomoell2018EUHFORIA}). However, despite a plethora of modeling efforts, accurately identifying CME arrival times remains challenging. Recent work by Kay et al. \cite{Kay2024CMEarrivals} showed that even state-of-the-art operational models still have an average CME arrival-time error of roughly 13 hours, with uncertainties in the background solar wind potentially remaining one of the major sources of forecasting error. Singh et al. \cite{Singh2023CMEarrivals} constrained the simulation ensemble members with the CME modeling and reduced the median absolute error (MAE) for the considered set of the CMEs from 8\,h to 4\,h. Nevertheless, the errors are still large and require a systematic assessment of their origin.

In this work, we build on a large grid of simulations produced with the GAMERA+Gibson-Low MHD model of the CME propagation in the inner heliosphere \cite{Provornikova2024GAMERAGL}. In the model, the solar wind background in region 0.1-1 au is driven by the WSA-ADAPT semi-empirical coronal model \cite{Arge:2000:WSA} and a CME is modeled by the analytical Gibson-Low model \cite{Gibson:1998:GL}.  The grid comprises 23{,}058 three-dimensional MHD simulations across three background solar wind states and a wide range of CME launch configurations. The simulation grid was generated within the Data-Optimized Interplanetary Coronal Mass Ejection Modeling (DOICMEM) project, and the present work builds on its results. 

The grid yields on the order of 10.5~million synthetic time series at L1, which makes it one of the largest physics-based datasets available both for studying CME propagation and for training machine learning models on it \cite{Sadykov2026COSPAR,Sadykov:2026:MLdataset}. We isolate the CME by subtracting an environment-matched background before any prediction is attempted, and then train a machine-learning regressor on the resulting labels. Beyond raw accuracy, the scale of the grid lets us address a question that is hard to study with the handful of labeled real events: how much of the predictable arrival-time signal comes from the CME's own launch properties, and how much from the ambient solar wind it propagates in?

\subsection{Hypothesis}
We hypothesize that CME arrival time at L1 is governed jointly by the CME's own kinematic and geometric properties (initial speed, angular width, and launch geometry relative to the observer) and by the state of the ambient solar wind it traverses, rather than by the launch parameters alone. We expect that (i) the transient CME signature can be isolated reliably by subtracting an adaptive, rotation-matched background template, (ii) once isolated, a multi-layer perceptron regressor can reproduce the simulated arrival times to within several hours, and (iii) machine learning models trained on several subsets of features will allow us to quantify contributions of different feature groups to the arrival time error. This set of hypotheses largely determines the structure of this paper.

\section{Methodology}

\subsection{Modeling Suite and 1\,AU Timeseries}

Two models feed into one another to build the grid of simulations in this work: the coronal magnetic field up to 21.5 $R_{Sun}$ is first reconstructed from solar photospheric magnetograms with the Wang--Sheeley--Arge (WSA) model \cite{Arge:2000:WSA,Arge:2004:WSA}, and its radial field $B_r$ and radial speed $v_r$ at 21.5 $R_{Sun}$ together with prescribed solar wind density and temperature serve as inner boundary conditions for GAMERA MHD model of the inner heliosphere \cite{Zhang:2019:GAMERA,Merkin:2016:GAMERA_a,Merkin:2016:GAMERA_b,Provornikova2024GAMERAGL}. GAMERA extends the flow from $0.1$ to 1\,$AU$ and is initially allowed to relax for about $10.5$ days of physical time, which yields a steady, rotation-only background wind. These backgrounds are drawn from three Carrington rotations representing different phases of the solar cycle: CR\,2095, CR\,2194, and CR\,2226 (rising, decaying, minimum activity, respecively).

A CME enters the inner heliosphere domain at the $21.5\,R_{Sun}$ boundary as an analytical Gibson--Low (GL) flux rope in self-similar expansion \cite{Gibson:1998:GL,Provornikova2024GAMERAGL}. That analytical form only seeds the initial CME structure: once past the boundary, the flux rope evolves under the full MHD equations, so its coupling to the background solar wind. The initial GL properties are listed in Table~\ref{tab:dataset}.

Outputs are archived as $1\,$AU shells. Because L1 sits only about $1\%$ of the Sun--Earth distance short of $1\,$AU, we use ``$1\,$AU shell'' and ``L1 shell'' to mean the same. A shell set follows each CME out to $120\,$h past launch, which is enough to contain even the slowest events (Figure~\ref{fig:cme_sim}). From every shell we extract synthetic L1 observers on a grid that steps roughly $4^\circ$ in longitude, offset in latitude as a function of longitude, and covers latitudes within about $8^\circ$ of the simulation equator at similar spacing. Summed over the grid, this leaves $\gtrsim$10.5 million time series. At each observer, the field and velocity vectors are rotated into the Geocentric Solar Ecliptic (GSE) frame standard for L1 data with $B_x$, $B_y$, $B_z$, $|B|$, $v_x$, $v_y$, $v_z$, $|v|$, density $\rho$, and pressure $p$. Every series is placed on a common $5$-minute grid of $90\,$h length, with the recording start shifted by launch speed so that slower CMEs begin later. The dataset is documented more fully in \cite{Sadykov2026COSPAR} (available on request) and in a paper in preparation \cite{Sadykov:2026:MLdataset}.


\begin{table*}[!tb]
\centering
\caption{Overview of the GAMERA-GL simulation dataset used in this study.}
\label{tab:dataset}
\scriptsize
\begin{tabular}{@{}p{0.30\linewidth} p{0.62\linewidth}@{}}
\toprule
\textbf{Property} & \textbf{Value} \\
\midrule
Complete MHD simulations & 23{,}058 \\
Synthetic L1 time series & $\sim$10.5 million \\
Background solar wind Carrington Rotations (CRs) & CR 2095, CR 2194, CR 2226 (rising, decaying, minimum solar activity) \\
Coronal / boundary model & ADAPT-WSA to $21.5\,R_{Sun}$ from photospheric magnetic field maps \\
Heliospheric MHD solver & GAMERA-GL until 1~AU using WSA-provided $B_{r}$ and $v_{r}$ at $21.5\,R_{Sun}$\\
CME model & Gibson--Low flux rope injected at $21.5\,R_{Sun}$ and evolving consistently with MHD equations \\
Varied CME (GL) parameters & CME front height speed (km/s): \{400, 500, 650, 900, 1500, 2500\}, CME angular width (degrees): \{10, 15, 20, 30, 60, 90, 120\}, launch latitude and longitude: 22 unique locations across three Carrington rotations, flux rope orientation (degrees): \{0, 60, 120, 180, 240, 300\}, flux rope chirality: \{left-handed, right-handed\}, flux rope maximum field strength (G): \{0.001, 0.002, 0.003, 0.004\}, flux rope topology-morphology parameter \cite{Provornikova2024GAMERAGL,Malanushenko:2020:CoNNs}: \{1.8, 2.5, 3.5, 4.1, 4.2, 5.0\} \\
L1 sampling resolution & $\sim$4.2$^{\circ}$ longitude $\times$ $\sim$3.9$^{\circ}$ latitude \\
Output cadence and duration & Interpolated to 300\,s temporal resolution; 90\,h in total \\
Recorded variables at 1~AU & $B_x, B_y, B_z, |B|, V_x, V_y, V_z, |V|$, density, pressure \\
\bottomrule
\end{tabular}
\end{table*}

\subsection{Background Isolation and Subtraction}

Our first objective is to separate the transient CME signature from the continuously evolving background solar wind. To achieve this, adaptive background templates are constructed from carefully selected simulations representing different solar cycle phases, namely, Carrington Rotation (CR) 2095, CR 2194, and CR 2226. Because the simulation times for which the 1\,AU shells were available varied depending on the CME initial speeds, and also because the relatively narrow CMEs did not affect the solar wind conditions in the direction opposite to their propagation, we have selected several representative simulations for each Carrington rotation which, together, provided the maximum possible span for the background for all locations at 1\,AU. 

In practice, for background derivation, we select the events with the leg-separation angle below $20^{\circ}$ (\texttt{GL\_legsang}), a launch speed under $1000~\mathrm{km\,s^{-1}}$ (\texttt{GL\_vel\_fh}), and a magnetic field strength parameter below $0.0015$ (\texttt{GL\_Bpar}), with candidate longitudes matched to within about $5^{\circ}$. Using three longitude seeds per rotation (used to construct the initial dataset) gave 27 background runs in total, nine each for CR 2095 and CR 2194, sampled on a fixed grid of observer points and stored as a compact per-rotation library. For the CR 2226 there were no, however, slow and narrow events allowing the full coverage; therefore, only two almost oppositely-directed CME cases were selected for each seed. This step resulted in the construction of background templates.

\begin{figure*}[!t]
\centering
\includegraphics[width=0.9\textwidth]{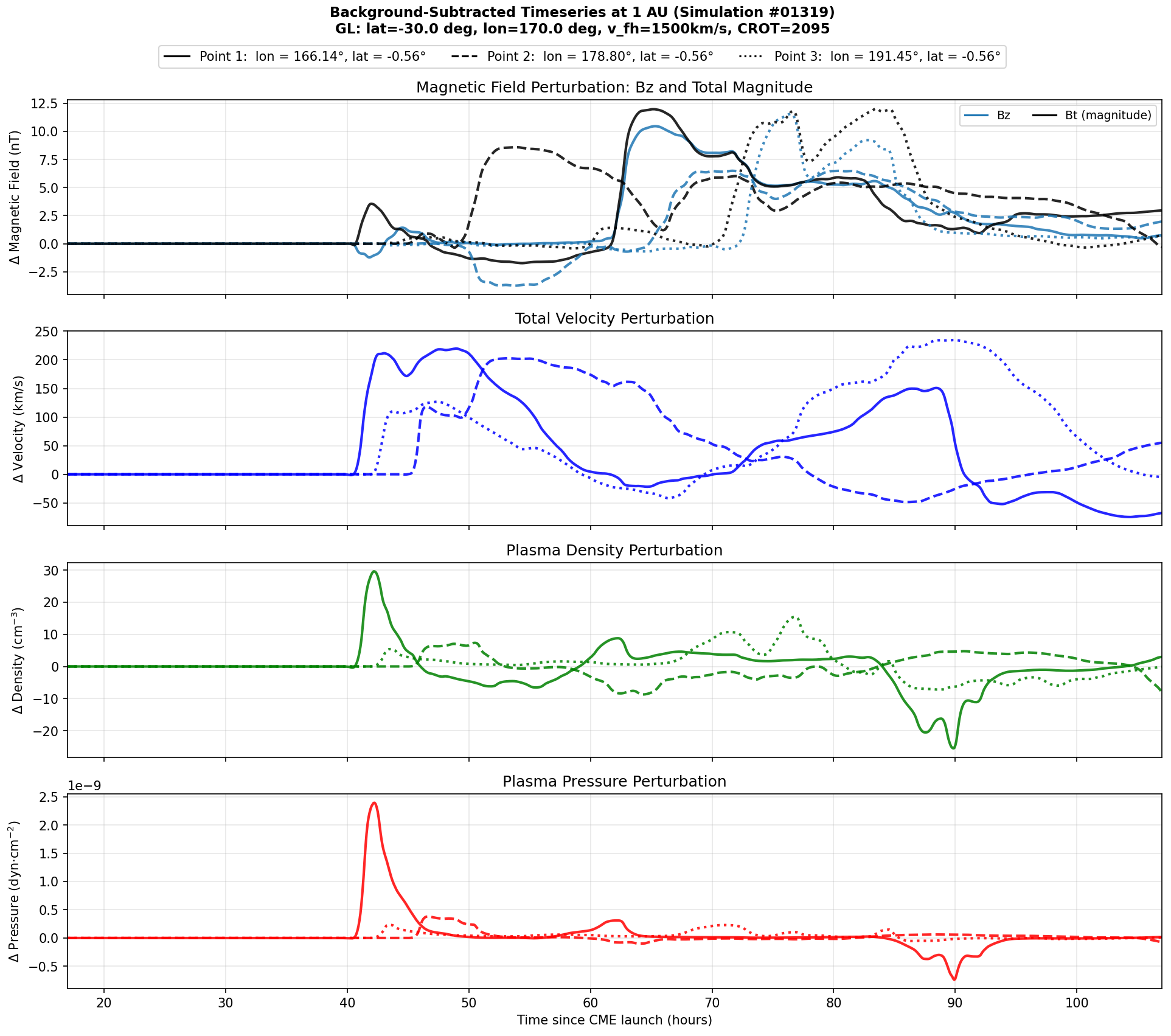}
\caption{Examples of the background-isolated timeseries at 1\,AU for the locations indicated in Figure~\ref{fig:cme_sim}. The panels, from top to bottom, represent the perturbations of the total magnetic field magnitude and its $B_{z}$ component in the GSE system, total plasma velocity, plasma density, and plasma pressure. Notice the near-zero perturbations before the CME impact indicating the robustness of the background subtraction.}
\label{fig:bgsub}
\end{figure*}

Because a CME run and its background were rarely recorded at exactly the same times, we interpolated the background onto the CME's own time grid (quadratically) before removing it, and we did the subtraction one parameter at a time:
\[ \Delta X(t) = X_{\mathrm{CME}}(t) - X_{\mathrm{background}}(t), \]
where $X$ is any of the ten L1 quantities. We carried this out for the three magnetic-field components and $|B|$, the three velocity components and $|V|$, the density, and the pressure. The illustration of the background subtraction is presented in Figure~\ref{fig:bgsub}, where it is notable that the signal is flat and near zero before $t\sim{}40$\,h, where the clear signature of the CME arrival is observed in one of the points.

\subsection{Recognition of the CME arrival moment}

Once the background is subtracted, what remains is dominated by the CME, and the task is to quantify the criteria for the CME arrival. An arrival is accepted only when several variables experience the disturbance relative to the background together. Concretely, the background-subtracted signal has to surpass the typical interpolation uncertainties for background subtraction. We have derived the maximum/peak interpolation errors for $|V|$ and $|B|$ and their components, as well as for the density and pressure, and require the CME to lead to perturbations at least twice larger than these values in order to be detected. The corresponding values are indicated in Table~\ref{tab:detection}. We note here that we require a simultaneous perturbation in one component associated with magnetic field measures, one component associated with velocities, density, and pressure, rather than considering them in isolation. To keep a single noisy spike from triggering a false detection, the disturbance must hold for at least 15 minutes, and we record the arrival as the onset of that sustained response. The 15 minute time has been motivated by the minimum time for the in-situ interplanetary CME detections by ARCANE algorithm \cite{Rudisser2026SpWea..2404537R}.

\begin{table}[!tb]
\centering
\caption{Adopted CME arrival detection criteria. Please note that the detection thresholds $|\Delta V|$ and $|\Delta B|$ apply to the magnitudes and individual components of these variables.}
\label{tab:detection}
\scriptsize
\begin{tabular}{@{}p{0.25\columnwidth} p{0.70\columnwidth}@{}}
\toprule
\textbf{Element} & \textbf{Description} \\
\midrule
Signal & Background-subtracted time series (all 10 parameters) \\
Detection thresholds & $|\Delta V| \ge 0.724\mathrm{km\,s^{-1}}$; $|\Delta n| \ge 0.443~\mathrm{cm^{-3}}$; $|\Delta B| \ge 0.182~\mathrm{nT}$; $|\Delta p| \ge 1.69\times10^{-11}~\mathrm{dyn\,cm^{-2}}$ \\
Coherence requirement & Consistent response across any of 4 magnetic field parameters ($B_x, B_y, B_z, |B|$), any of velocity parameters ($V_x, V_y, V_z, |V|$), density, and pressure \\
Hold time & 15 minutes (to reject single-point spikes) \\
Arrival definition & First multi-parameter response that is held \\
\bottomrule
\end{tabular}
\end{table}

\begin{figure*}[!tb]
\centering
\includegraphics[width=0.49\linewidth]{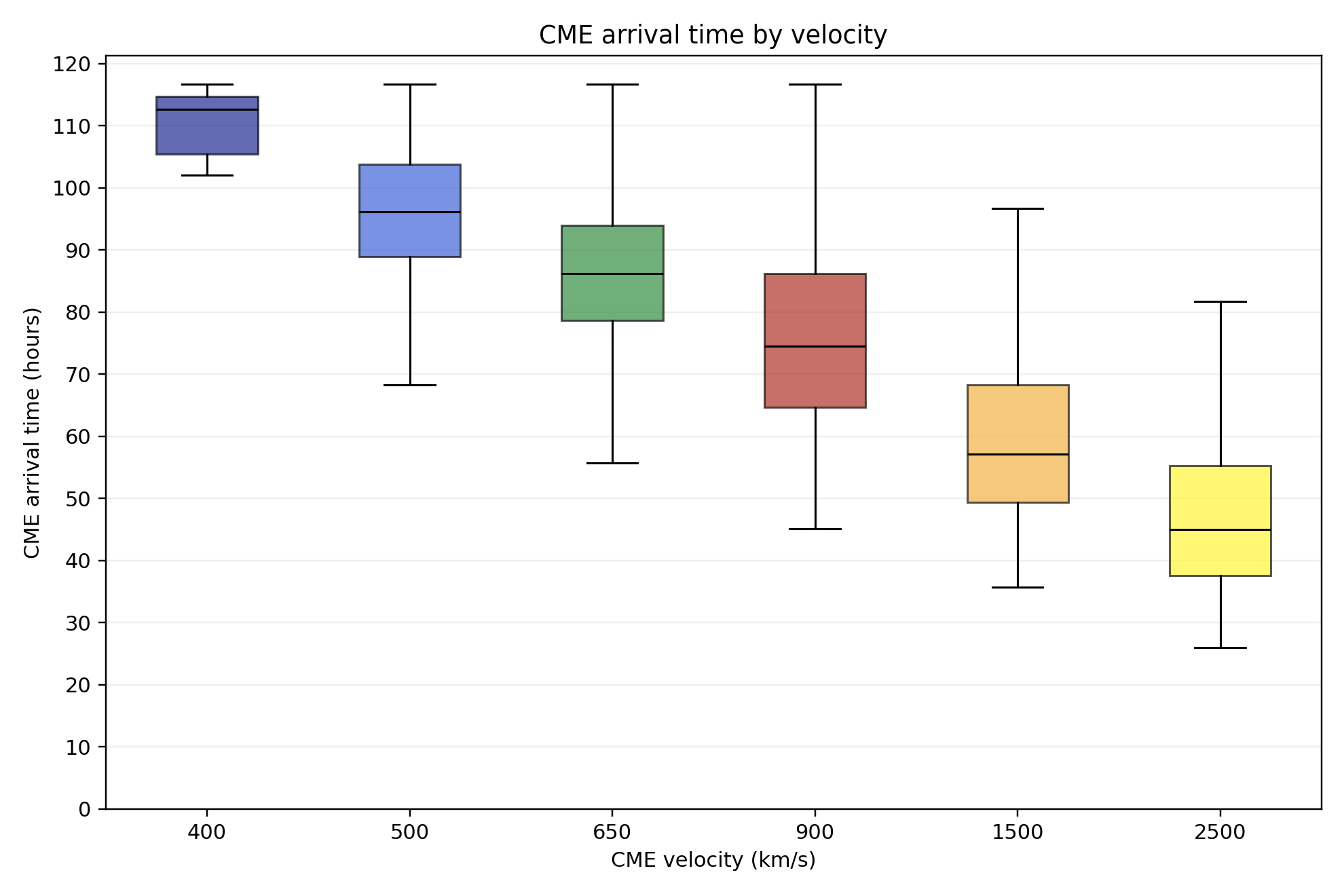}
\includegraphics[width=0.49\linewidth]{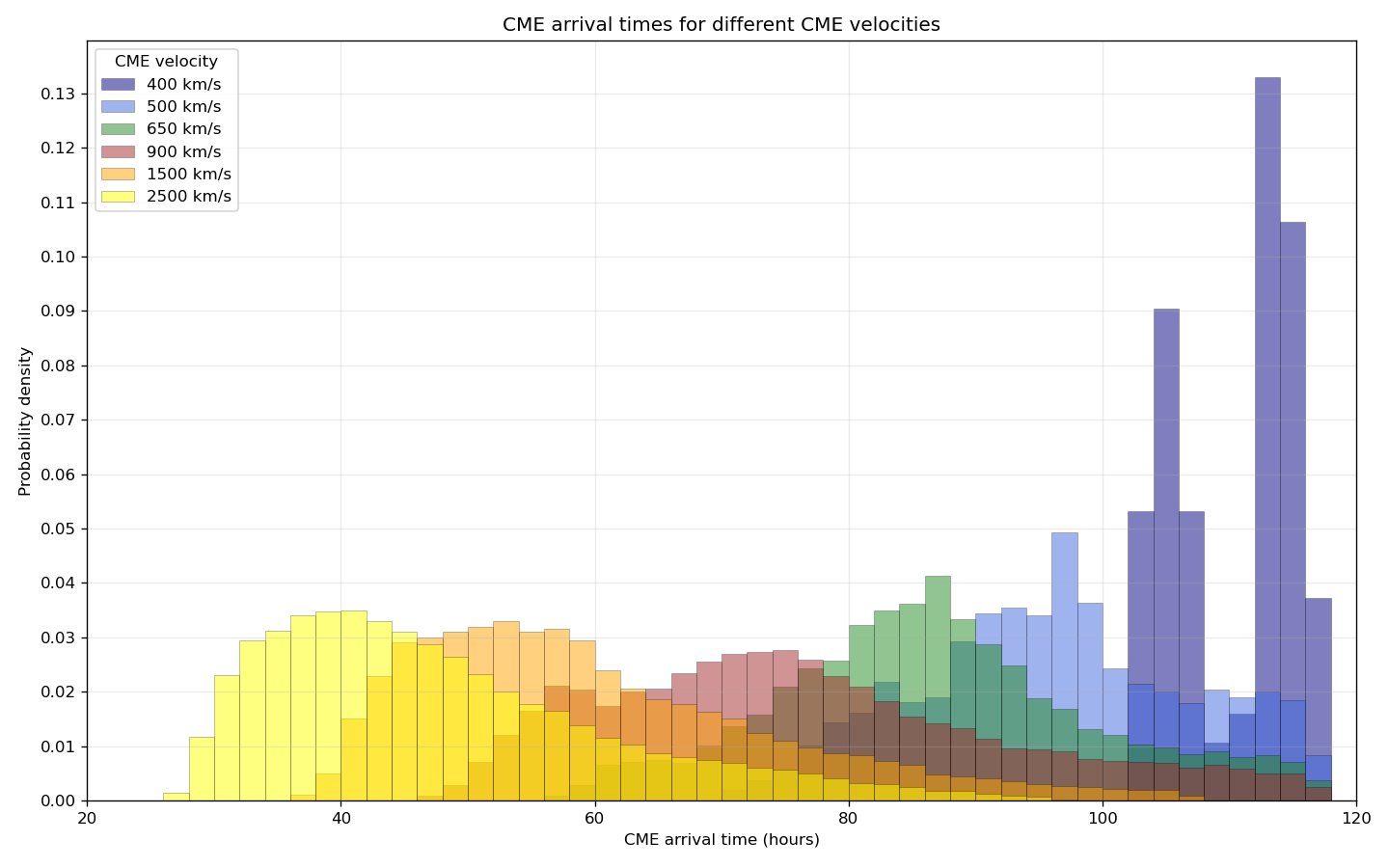}
\caption{Left: distribution of the CME arrival times for each value of the CME front height velocity parameter. Boxes indicate the interquartile range and the horizontal line inside each box marks the median. Whiskers extend to the most extreme data points. Right: same arrival time distributions presented as the probability density functions.}
\label{fig:arrivaltimes_dist}
\end{figure*}

With arrivals in hand, we combine them with the CME's own properties, such as launch speed, launch longitude, and the longitudinal separation between the CME and the observing point, and grouped the events by initial speed (400, 500, 650, 900, 1500, and $2500~\mathrm{km\,s^{-1}}$) so that the speed dependence of the transit time could be understood directly, and displayed in Figure~\ref{fig:arrivaltimes_dist}. The right panel here indicates the expected distributions for each velocity with the extended tails for high-speed CMEs (most probably, CME side impacts). The bi-modal nature of the distribution for the 400\,km/s arises from the limited number of cases where the impact of such a slow CME has been detected (only 94 arrival detections). As a sanity check we also computed a purely magnetic arrival, the first time $|B|$ and its components reach $1~\mathrm{nT}$ (a threshold comparable, by order of magnitude, to the background solar wind magnetic fields in the simulations); we note here that this and other implementations of arrival time detections do not result in the significant changes of the dataset. The statistics of arrival times and related simulation numbers is presented in Table~\ref{tab:velstats}.

\begin{table}[!tb]
\centering
\caption{L1 samples with a detected CME arrival and number of unique simulation runs for each CME launch velocity.}
\label{tab:velstats}
\scriptsize
\setlength{\tabcolsep}{4pt}
\begin{tabular}{@{}r r r r@{}}
\toprule
\textbf{Velocity (km/s)} & \textbf{Samples} & \textbf{Share (\%)} & \textbf{Unique runs} \\
\midrule
400  & 94        & 0.01   & 35     \\
500  & 12,152    & 0.82   & 1,168  \\
650  & 46,215    & 3.11   & 1,328  \\
900  & 290,789   & 19.54  & 4,473  \\
1500 & 409,240   & 27.50  & 5,161  \\
2500 & 729,392   & 49.02  & 6,498  \\
\midrule
All  & 1,487,882 & 100.00 & 18,663 \\
\bottomrule
\end{tabular}
\end{table}

\subsection{ML Model for Arrival Time Prediction}

The dataset points with detected CME arrival times comprise 1{,}487{,}882 time-series samples drawn from 18{,}663 distinct GAMERA-GL runs (see Table~\ref{tab:velstats}).This number is smaller than the original number of 23{,}058 simulation runs because not every run has caused CME-related perturbations at the hypothetical L1 locations (in particular, relatively narrow CMEs launched at high latitudes missed them). Using the physics-based arrival labels described above, we train a feed-forward neural-network regressor to predict the CME arrival time (target \texttt{cme\_time}) directly from the simulation inputs. To prevent the network from memorizing the absolute launch geometry, we replace the CME launch latitude by its value relative to the observer, $\mathrm{GL\_lat\_rel} = \mathrm{GL\_lat} - \mathrm{lat\_L1}$, while retaining \texttt{lat\_L1} itself as a separate input. We also utilize $\mathrm{GL\_lon\_rel} = \mathrm{GL\_lon} - \mathrm{lon\_L1}$. Without these transformations the network tends to lock onto the small set of discrete \texttt{GL\_lat} and \texttt{GL\_lon} values present in the simulation grid (only 22 unique locations across three boundary conditions) and uses them effectively as a lookup key. The relative coordinate removes the shortcut while preserving the physically meaningful CME--observer geometry.

Each sample is described by up to 83 features in three families: (i)~eight Gibson--Low CME parameters, split into \emph{kinematic/geometric} properties, namely launch front speed (\texttt{GL\_vel\_fh}), angular width (\texttt{GL\_legsang}), relative launch latitude (\texttt{GL\_lat\_rel}) and relative longitude (\texttt{GL\_lon\_rel}), and \emph{magnetic/structural} properties, namely field-strength parameter (\texttt{GL\_Bpar}), orientation (\texttt{GL\_orientation}), chirality (\texttt{GL\_CHIRAL}), and topology-morphology (\texttt{GL\_topmorph}); (ii)~72 spherical-harmonic boundary-condition coefficients describing the WSA background solar wind, with 36 radial-magnetic-field ($b_r$) and 36 radial-speed ($v_r$) coefficients, each with cosine and sine terms for degree and order $\ell, m = 0\ldots5$; and (iii)~three observer and environment descriptors, namely the CME density (\texttt{Den\_CME}), the observer latitude at L1 (\texttt{lat\_L1}), and the Carrington-rotation identifier (\texttt{CROT}).

We consider a fully connected neural network regressor (scikit-learn \texttt{MLPRegressor}) with $\tanh$ activations, trained with the Adam optimizer at a fixed learning rate $lr = 0.001$ and a batch size of 256 on standardized input features (using Z-score standardization). The hyperparameters are selected by a grid search over the network architecture and the $L_2$ regularization parameter $\alpha$. The architecture grid consists of 11 configurations of hidden-layer sizes: (16), (32), (64), (128), (32, 16), (64, 32), (128, 64), (64, 32, 16), (128, 64, 32), (128, 64, 64), and (128, 128, 64). The regularization grid is $\alpha \in \{10^{-3}, 3.16\times{}10^{-3}, 10^{-2}, 3.16\times{}10^{-2}, 10^{-1}, 3.16\times{}10^{-1}, 1\}$. The network is trained epoch by epoch, and after each epoch the RMSE is evaluated on a validation set held out from the training data. Training stops when the validation RMSE has not improved for 2 consecutive epochs. 

In order to avoid within-simulation data leakage, both the train-test and sub-train-validation splits were done in such a way that the timeseries from one simulation are located within one dataset (hereafter run-level splits). 20\% of the data were held out for testing, and 20\% of the training dataset for early stopping. The best configuration for each feature mode is the one with the lowest validation RMSE averaged over the 10 repetitions initialized with different random seeds.

However, it has been realized that even the run-level split can potentially produce memorization issues and lead to overfitting. The possible reason for this behavior is that certain properties of the CME magnetic fields (such as magnetic field chirality, \texttt{GL\_CHIRAL}) do not impact the CME dynamics much. Therefore, we have performed another split strategy, the group split based on the combined launch location, CME speed, and CME width. This split results in 254 unique cases, which provide sufficient diversity for a later train-validation-test split.

To quantify how much of the arrival-time skill comes from the CME launch parameters versus the ambient solar wind, we retrain the network from scratch in four nested feature modes: (1)~\emph{all inputs} (83 features); (2)~\emph{no
boundary}, dropping the 72 $b_r$/$v_r$ harmonics and keeping only the 11 CME and observer features; (3)~\emph{no magnetic}, dropping the four magnetic/structural GL parameters (\texttt{GL\_Bpar}, \texttt{GL\_orientation}, \texttt{GL\_CHIRAL}, \texttt{GL\_topmorph}) while keeping the boundary condition spherical harmonics; and (4)~\emph{kinematic-only}, keeping only the CME kinematic/geometric parameters, the density, and the observer descriptors (seven features, with no magnetic and no boundary inputs). The optimal network configurations and all results are presented in Table~\ref{tab:ml}.

\section{Results}

\begin{figure*}[!tb]
\centering
\includegraphics[width=0.49\linewidth]{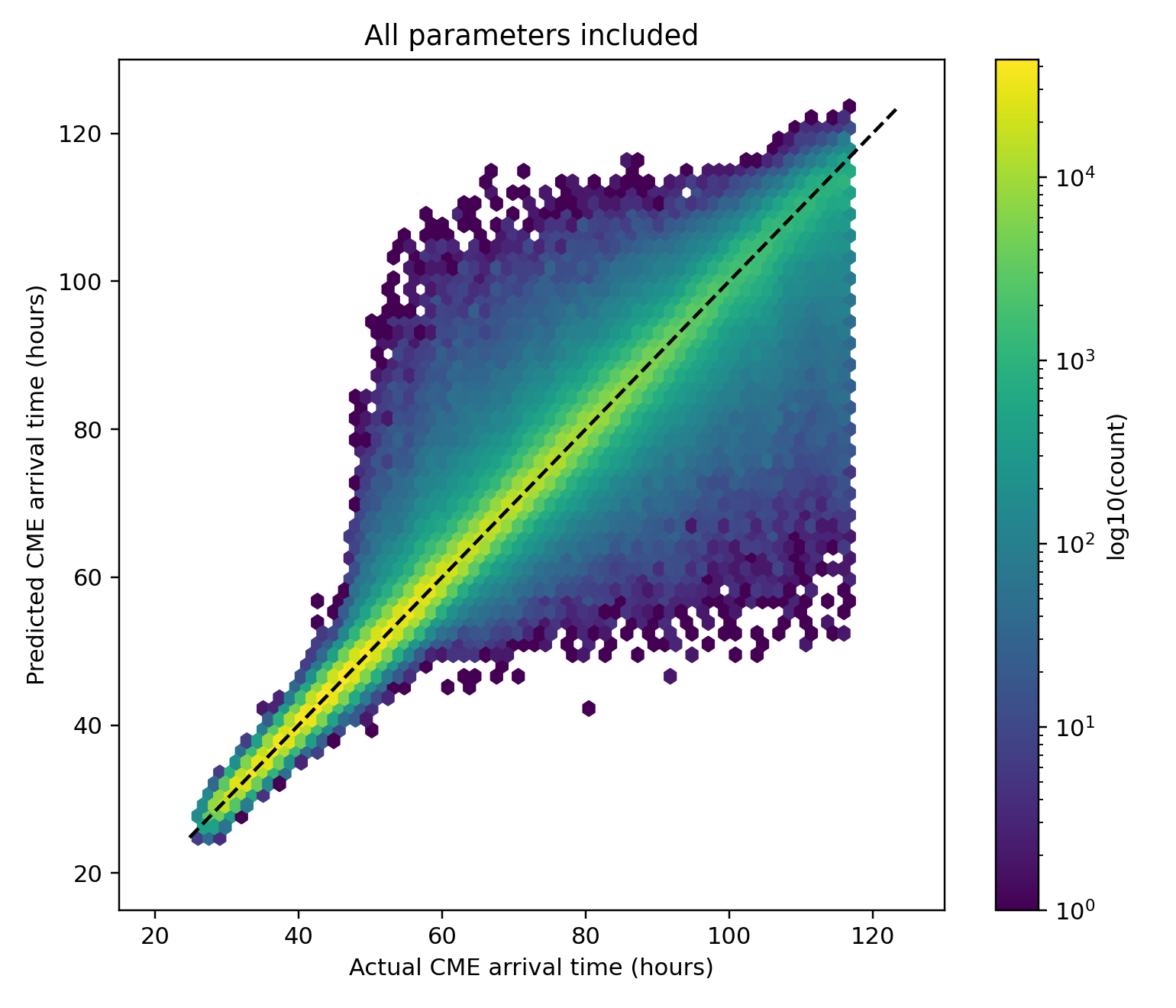}
\includegraphics[width=0.49\linewidth]{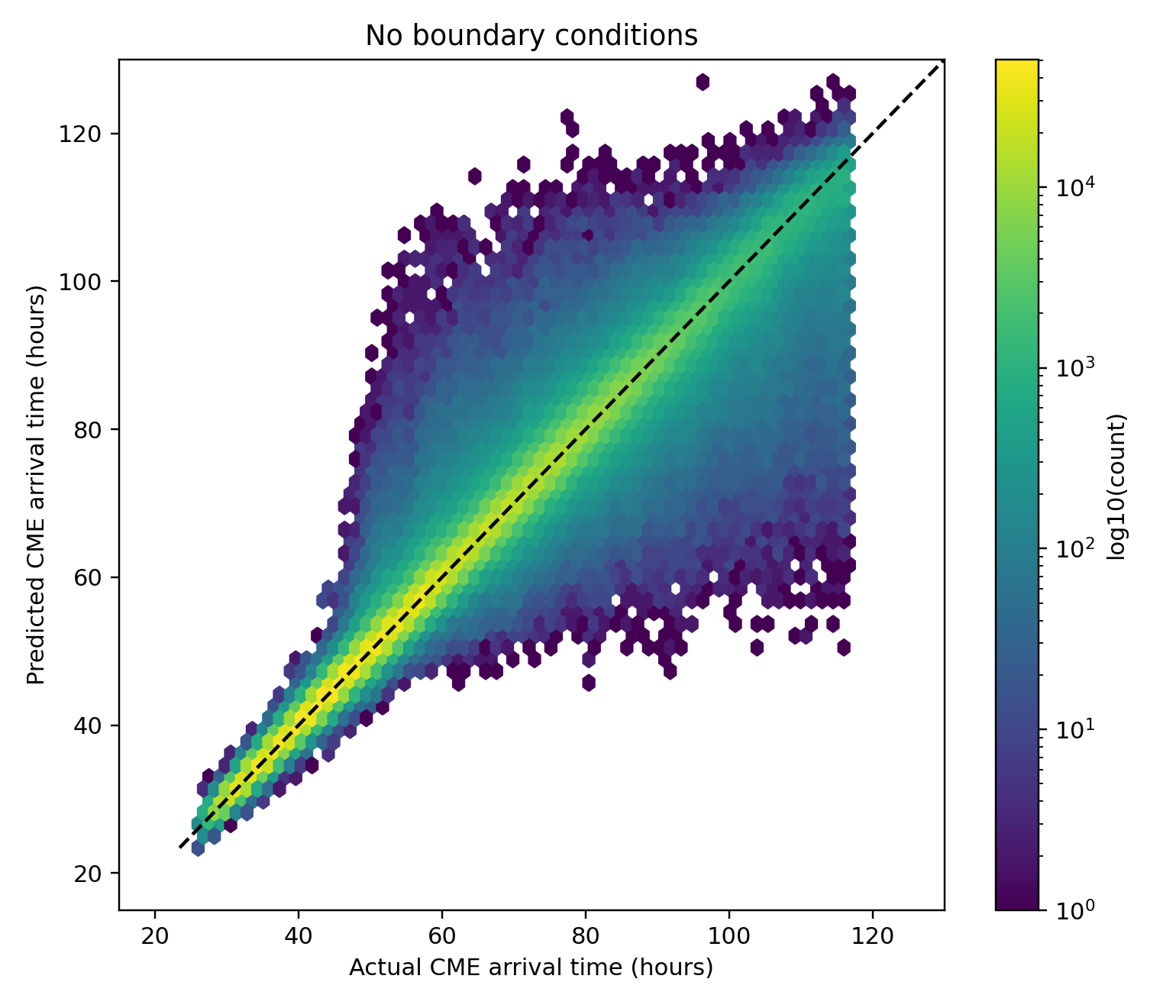} \\
\includegraphics[width=0.49\linewidth]{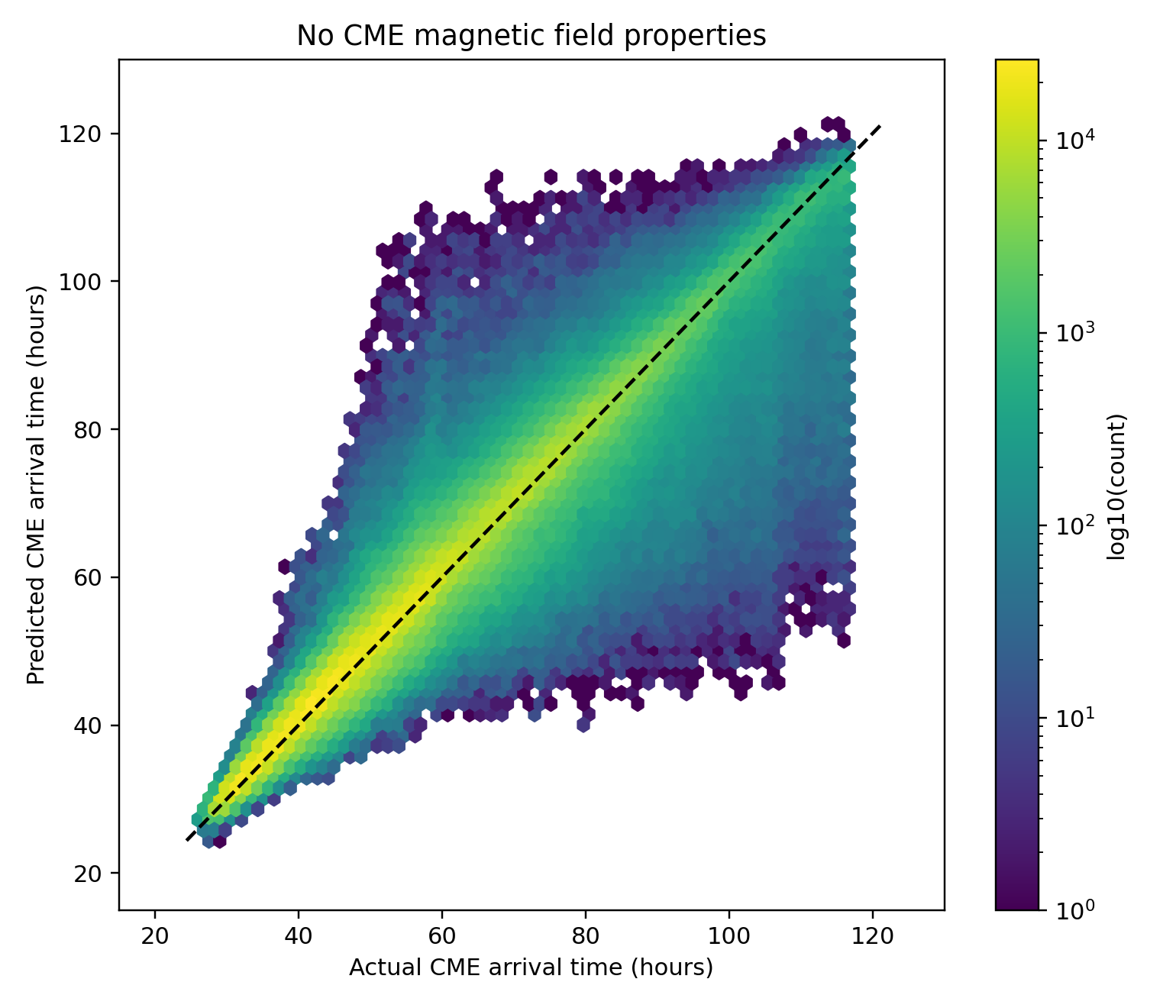}
\includegraphics[width=0.49\linewidth]{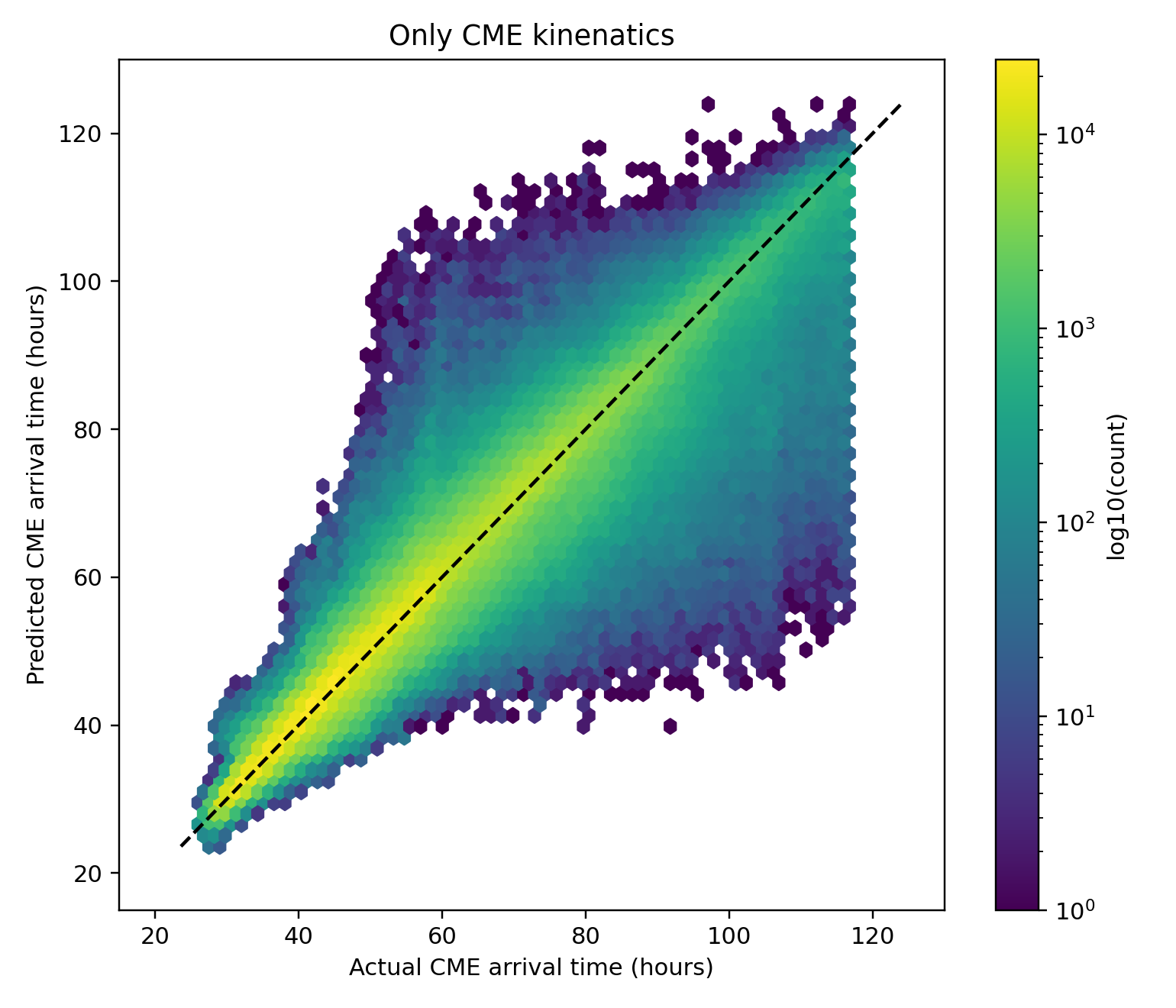}
\caption{Hex-bin scatterplots of the CME arrival times (actual VS predicted) for four different settings: all information is considered (top left), no boundary condition information is included (top right), no CME magnetic field properties are included (bottom left), and only CME kinematic properties (i.e., no magnetic field and boundary condition information) are included (bottom right). The dashed line indicates one-to-one relation corresponding to the ideal prediction.}
\label{fig:arrivaltimes_modes_nogrouping}
\end{figure*}

Figure~\ref{fig:arrivaltimes_modes_nogrouping} illustrates the hex-bin scatter plot of the predicted versus the actual arrival times. Different panels correspond to different feature groups considered. As evident, the predictions are the most accurate for early arrival times ($\lesssim$45\,h), which is, in principle, possible only for the fast CMEs (with $v\in[1500, 2500]~\mathrm{km\,s^{-1}}$). The modes with all inputs and with no boundary conditions included look very similar, with a slightly wider spread (up to individual prediction points) for the case with no boundary conditions. Interestingly, the removal of magnetic field properties leads to a visibly broader distribution, showing larger disagreement. The same is true, with an even wider spread, for the kinematics-only CME properties.

The cumulative metrics for predictions are summarized in Table~\ref{tab:ml}. As evident here, for the run-level split, the prediction RMSE (3.09$\pm$0.07\,h) and MAE (1.67$\pm$0.04) measures remain almost the same if the boundary conditions are excluded. In fact, the absence of boundary conditions results in more samples with arrival-time deviations of less than 30 minutes and 1 hour from the targets. However, the predictions are significantly worse for the cases with no magnetic features (RMSE 5.14$\pm$0.07\,h) and with only CME kinematics included (RMSE 5.63$\pm$0.08\,h). This is physically justifiable: the magnetic field strength affects the expansion of the CME and its interaction with the ambient solar wind, and the topology and orientation of the flux rope determine the CME front, overall affecting arrival times. Nevertheless, the metrics indicating essentially the same performance with and without boundary conditions pose questions. Does that mean that the 72 spherical harmonic coefficients carry little information beyond identifying the Carrington rotation? Or does that mean that the model memorizes the runs and applies the memorized predictions to similar runs in the test dataset with slightly deviating magnetic field properties?

\begin{table*}[!tb]
\centering
\caption{Arrival time prediction performance of the neural network on the test sets ($\sim$20\% of the data, 10 folds) for the four feature modes. For MAE, the median $\pm$ median absolute deviation (MAD) over the runs is presented. Two partitioning scenarios are shown. In the run-level split, all L1 samples from a single simulation run fall into the same partition. In the group split, all runs sharing the same CME launch location, initial speed, and width (254 unique groups) fall into the same partition.}
\label{tab:ml}
\scriptsize
\setlength{\tabcolsep}{4pt}
\begin{tabular}{@{}l c c c c c c c c@{}}
\toprule
\textbf{Feature mode} & \textbf{Architecture} & \textbf{Reg. $\alpha$} & \textbf{N Feat.} & \textbf{RMSE (hr)} & \textbf{MAE (hr)} & \textbf{$R^2$} & \textbf{$\le$30 min (\%)} & \textbf{$\le$1 hr (\%)} \\
\midrule
\multicolumn{7}{@{}l}{\textit{Run-level split}} \\
All inputs                  & (128, 128, 64) & 0.001 & 83 & 3.09$\pm$0.07 & 1.67$\pm$0.04 & 0.973$\pm$0.001 & 27.2$\pm$1.1 & 49.9$\pm$1.6 \\
No boundary (keep magnetic) & (128, 128, 64) & 0.001 & 11 & 3.11$\pm$0.07 & 1.68$\pm$0.02 & 0.973$\pm$0.001 & 28.3$\pm$1.4 & 51.0$\pm$1.8 \\
No magnetic (keep boundary) & (128, 128, 64) & 0.001 & 79 & 5.14$\pm$0.07 & 3.21$\pm$0.08 & 0.926$\pm$0.002 & 14.5$\pm$0.9 & 28.0$\pm$1.6 \\
Kinematic only              & (128, 128, 64) & 0.001 & 7  & 5.63$\pm$0.08 & 3.63$\pm$0.07 & 0.911$\pm$0.002 & 13.0$\pm$1.1 & 25.1$\pm$1.9 \\
\midrule
\multicolumn{7}{@{}l}{\textit{Group split (launch location, CME speed, CME width)}} \\
All inputs                  & (128, 128, 64) & 0.100 & 83 & 6.47$\pm$0.48 & 4.30$\pm$0.28 & 0.876$\pm$0.016 & 10.1$\pm$1.7 & 19.9$\pm$3.0 \\
No boundary (keep magnetic) & (128, 128, 64) & 0.032 & 11 & 7.02$\pm$0.79 & 5.14$\pm$0.53 & 0.854$\pm$0.029 & 8.4$\pm$1.7  & 16.6$\pm$3.1 \\
No magnetic (keep boundary) & (128, 64) & 0.316 & 79 & 8.02$\pm$0.54 & 5.72$\pm$0.26 & 0.810$\pm$0.022 & 7.3$\pm$0.9  & 14.5$\pm$1.8 \\
Kinematic only              & (128, 128, 64) & 0.100 & 7  & 9.64$\pm$0.90 & 7.37$\pm$0.51 & 0.725$\pm$0.045 & 4.9$\pm$0.7  & 9.9$\pm$1.4  \\
\bottomrule
\end{tabular}
\end{table*}

\begin{figure*}[!tb]
\centering
\includegraphics[width=0.49\linewidth]{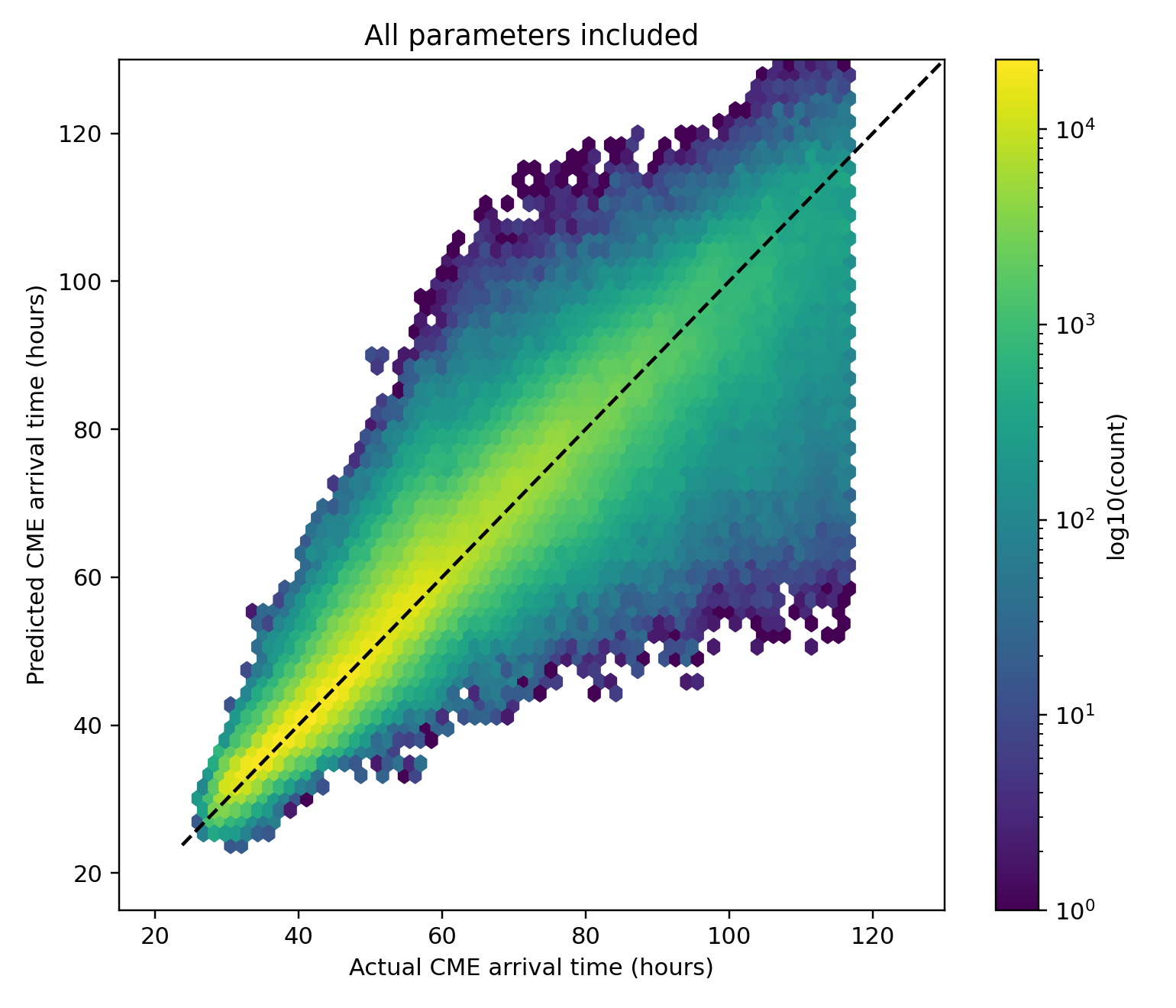}
\includegraphics[width=0.49\linewidth]{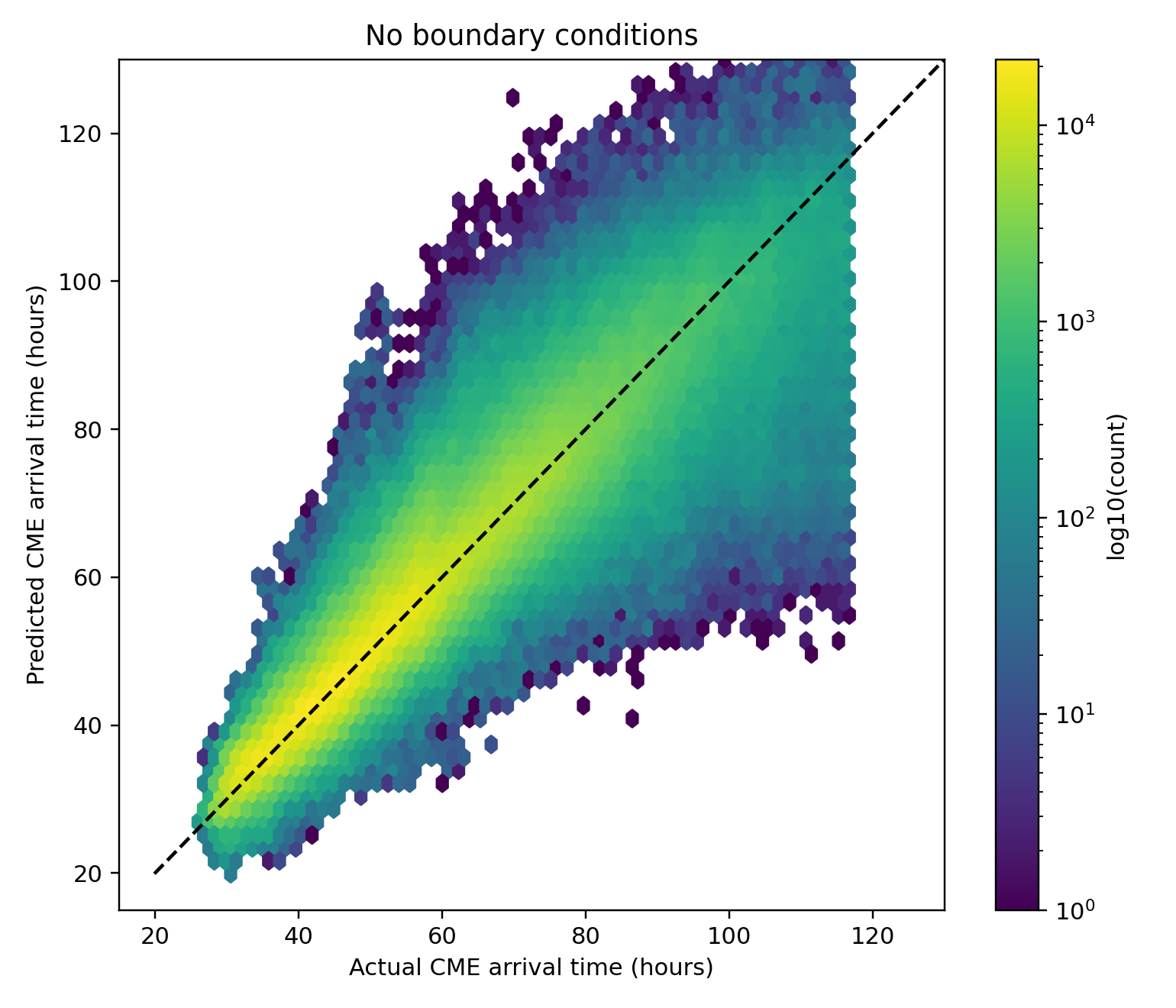} \\
\includegraphics[width=0.49\linewidth]{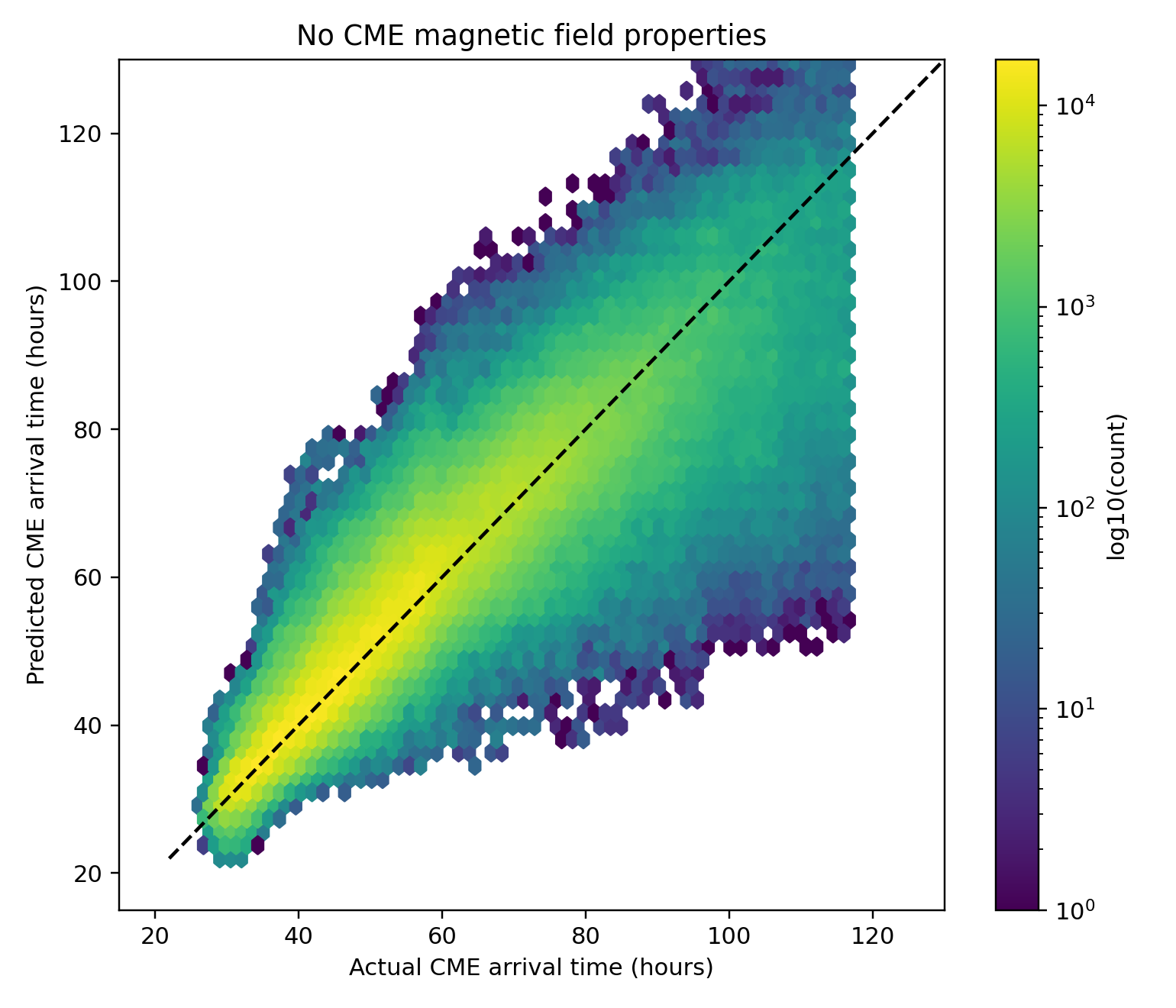}
\includegraphics[width=0.49\linewidth]{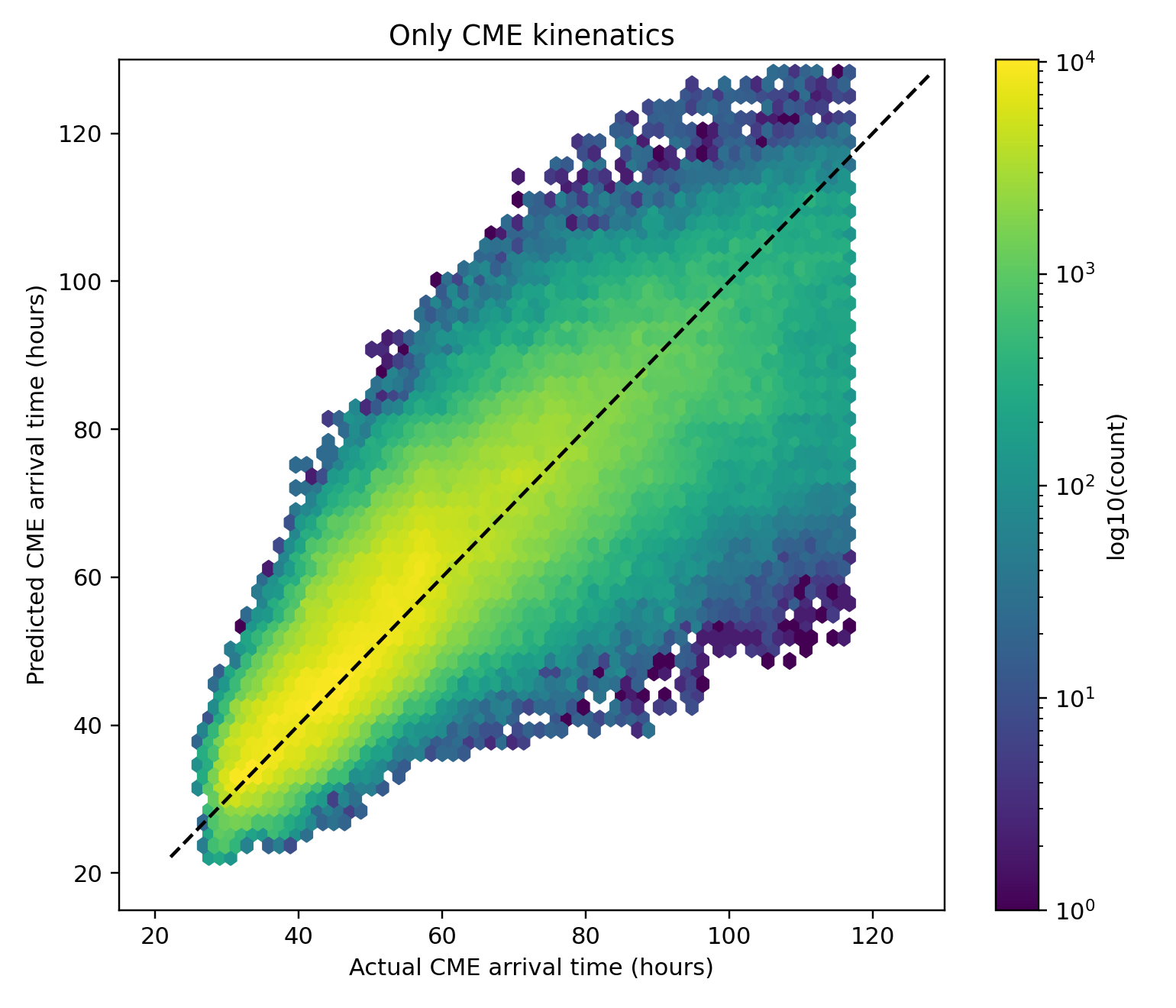}
\caption{Same as in Figure~\ref{fig:arrivaltimes_modes_nogrouping} but with ensuring the CME grouping based on location, speed, and width.}
\label{fig:arrivaltimes_modes_withgrouping}
\end{figure*}

To answer this question, we perform the grouping of simulations based on the CME launch location, its speed, and its width (which means that all CME simulations with the unique combination of these parameters are either in the train, or in the validation, or in the test dataset). The results are illustrated in Figure~\ref{fig:arrivaltimes_modes_withgrouping} and also presented in Table~\ref{tab:ml}. It is evident that the scatterplots are significantly wider than those presented in Figure~\ref{fig:arrivaltimes_modes_nogrouping}, yet still show that the data are concentrated near the one-to-one line. For the full set of input features, the RMSE increases about twice, from 3.09$\pm$0.07 to 6.47$\pm$0.48\,h. The MAE is 4.30$\pm$0.28\,h and $R^2=0.876$. The fraction of predictions within 1\,h drops to 19.9\%, and within 30\,min is only 10.1\%. It is notable that the uncertainties increase as well, to 0.5--0.9\,h in RMSE, indicating that the performance depends more strongly on the particular split scenario. At the model selection level, the grid search favors considerably stronger regularization than in the run-level split, with $\alpha=0.03$--$0.3$ instead of $10^{-3}$. Overall, we feel that the group split provides a more realistic estimate of how well the models generalize to CME configurations not seen during training.

Regarding the ranking of feature modes, the CME magnetic parameters remain the most important group of features after kinematics: without the magnetic field parameters, the RMSE increases by $\sim$24\% (8.02\,h), and if kinematics only is considered, by $\sim$49\% (9.64\,h, $R^2=0.725$). Removing the boundary conditions increases the error by only $\sim$8\% (7.02\,h), a difference comparable to the fold-to-fold scatter. We note, though, that the background solar wind information becomes especially useful in the absence of the CME magnetic field properties: with the boundary conditions retained the RMSE is 8.02$\pm$0.54\,h, compared to 9.64$\pm$0.90\,h for the kinematic-only case without them.

\begin{figure*}[!htb]
\centering
\includegraphics[width=\linewidth]{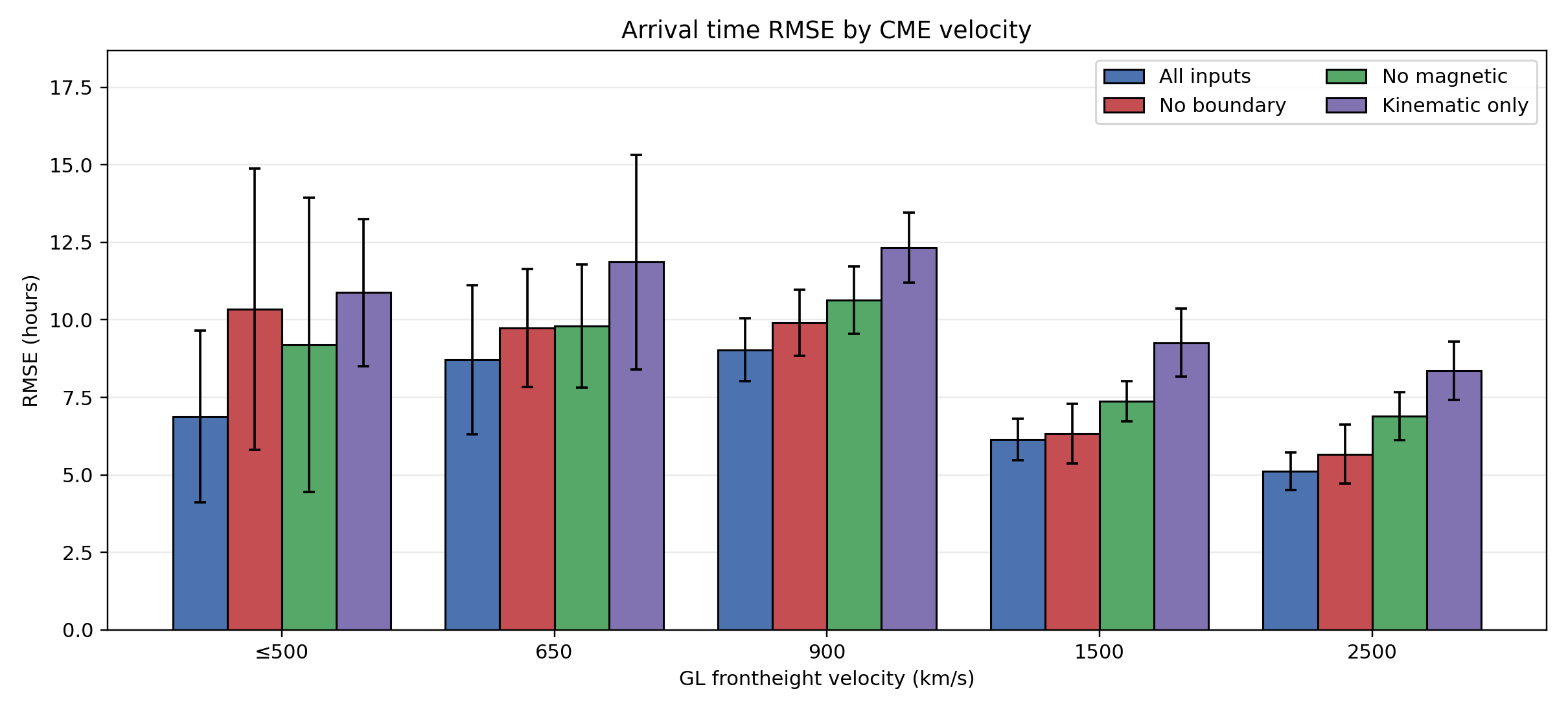}
\caption{Test set arrival time RMSEs broken down separately for each case of the CME launch velocity, for the four feature modes. The same information is presented in Table~\ref{tab:rmsevel} as well. The cases of 400\,km/s and 500\,km/s CMEs are combined into a single bin due to the sparse representation of arrival times for CMEs with 400\,km/s}
\label{fig:rmsevel}
\end{figure*}

To further investigate the arrival time prediction for the realistic group-split case, we display the prediction RMSE scores in Figure~\ref{fig:rmsevel}, and list them in Table~\ref{tab:rmsevel}. We see several interesting dependencies here. First, for CMEs launched at 900\,km/s and faster, the error decreases with increasing speed (which is likely the result of the smaller number of cases of slower CMEs, see Table~\ref{tab:velstats}). With all inputs, the RMSE is 9.03\,h at 900\,km/s, 6.14\,h at 1500\,km/s, and 5.11\,h at 2500\,km/s. The fastest CMEs travel to 1\,AU in the shortest time, so there is less time for interaction with the ambient solar wind to accumulate deviations from the initial kinematics. For the fast CMEs, the ranking of the feature modes is also consistent, with relatively small validation-fold uncertainties. Another interesting dependency is that the relative importance of the boundary conditions and the CME magnetic field parameters inverts for slow ($v_{CME}\lesssim650$\,km/s) CMEs, with the boundary conditions playing a more important role. For example, for $v_{CME}\lesssim{}500$\,km/s cases, the absence of boundary conditions increases the RMSE from 6.88$\pm$2.77\,h to 10.35$\pm$4.54\,h. This is overall consistent with the physical picture of slow CMEs spending more time in the ambient solar wind, interacting with it more, and therefore deviating from the expected arrival times. Nevertheless, we point here that the slow CMEs ($v_{CME}\lesssim{}500$\,km/s) constituted only $\sim$0.83\% of the dataset. Further work is needed to understand how well the neural network generalizes to this population and to what extent the large arrival-time uncertainties for slow CMEs can be explained by a potentially poor generalization.

\begin{table}[!tb]
\centering
\caption{Test-set RMSE (hours) for each CME front height velocity option and all four feature modes. The 400 and 500~km/s launch velocities are merged into a single bin because of not many cases presented for the 400\,km/s velocity CMEs.}
\label{tab:rmsevel}
\scriptsize
\setlength{\tabcolsep}{2pt}
\begin{tabular}{@{}l c c c c c@{}}
\toprule
\textbf{Velocity (km/s)} & \textbf{$\leq$500} & \textbf{650} & \textbf{900} & \textbf{1500} & \textbf{2500} \\
\midrule
All inputs      & 6.88$\pm$2.77  & 8.71$\pm$2.40  & 9.03$\pm$1.01  & 6.14$\pm$0.67 & 5.11$\pm$0.60 \\
No boundary     & 10.35$\pm$4.54 & 9.74$\pm$1.90  & 9.90$\pm$1.07  & 6.33$\pm$0.97 & 5.67$\pm$0.96 \\
No magnetic     & 9.19$\pm$4.75  & 9.80$\pm$1.99  & 10.64$\pm$1.09 & 7.37$\pm$0.64 & 6.89$\pm$0.77 \\
Kinematic only  & 10.88$\pm$2.37 & 11.86$\pm$3.46 & 12.33$\pm$1.12 & 9.26$\pm$1.10 & 8.36$\pm$0.93 \\
\bottomrule
\end{tabular}
\end{table}

\section{Summary and Conclusions}

In this work, we utilize the large grid of GAMERA-GL MHD simulations of the CME in order to understand how various CME properties affect their arrival times. The key summary points are as follows:
\begin{itemize}
    \item Approximately 23{,}000 GAMERA-GL simulations initialized with three background solar wind configurations were sampled at hypothetical L1 locations, resulting in $\sim$10.5 million synthetic time series. After background subtraction and a multi-parameter arrival criterion sustained for at least 15\,min, 1.49 million samples from 18{,}663 runs contain detected CME arrivals.
    \item A neural network regressor is tuned over the architecture and the $L_2$ regularization strength with early stopping, and trained in four feature-ablation modes (with groups of features corresponding to boundary conditions and CME magnetic field properties removed).
    \item The run-level split leads to the strong prediction accuracy. With all inputs, the RMSE is 3.09 \, h, $R^2=0.973$, and $\sim$50\% of predictions fall within 1 \, h of the actual arrival time. Removing the boundary conditions does not change the performance, whereas removing the CME magnetic parameters increases the RMSE to 5.14\,h. This raises the question of whether the network memorizes similar runs.
    \item The group split leads to a more realistic picture. Grouping the runs by launch location, speed, and width (254 groups) doubles the error, to an RMSE of 6.47\,h with $R^2=0.876$ and $\sim$20\% of predictions within 1\,h, and requires stronger regularization. The importance ranking is preserved: the CME magnetic parameters are the most important features after kinematics (+24\% RMSE without them, +49\% with kinematics only), whereas the boundary conditions provide a modest improvement, comparable to the fold-to-fold scatter.
    \item Prediction errors depend on CME speed. The fastest CMEs are predicted best, with RMSE of $\sim$5\,h at 2500\,km/s compared to $\sim$9\,h at 900\,km/s. For slow CMEs, the boundary conditions appear to be relatively more important, consistent with their longer interaction with the ambient solar wind, although the uncertainties for these CMEs are large.
\end{itemize}

To conclude, the CME arrival time depends on both the CME launch properties and the ambient solar wind, but not equally: the CME kinematic and magnetic parameters carry most of the predictable signal, while the background solar wind provides a secondary contribution to the uncertainties at the level of $\sim$0.5\,h of the additional arrival time error. Nevertheless, it is most valuable for slow CMEs and when the CME magnetic information is unavailable (its absence increases the arrival time error by $\sim$1.5\,h in this case). Incorrect or absent assumption about magnetic field properties of the CMEs can lead to an additional arrival time error of $\sim$1.5\,h as well in case the boundary conditions are known. Since the near-Sun kinematic CME properties are exactly the quantities a coronagraph can measure before impact, and the magnetic properties of the CMEs are generally not known, these results support the importance of the background solar wind for the CME arrival time prediction. 

Overall, our results should be viewed as an assessment of which properties of the CMEs, if measured at all and more accurately, would help reducing arrival time errors, rather than as an operational forecasting tool. Still, the primary error (RMSE or 6.47$\pm$0.48, MAE 4.30$\pm$0.28) is likely a result of a complex interaction of the CME with the background solar wind that the limited spherical harmonic expansion (up to $L=5$ only) cannot capture. The other limitations of current work related to the parameter space sampling (see Table~\ref{tab:dataset}) and limited boundary condition cases (three Carrington rotations); more comprehensive ML models can also be useful to further reduce the modeled arrival time errors. Future works can address these limitations.

\section*{Acknowledgment}
This project has been supported by the NASA Living With the Star (LWS) program under award \#80NSSC24K1111 and GSU Physics\&Astronomy summer research fellowship. We acknowledge NASA HECC, GSU DMLab, and GSU ARCTIC, for the use of computational resources. AI-based tools (a large-language-model coding and writing assistant) were used to help implement some portions of the analysis code and to edit portions of the manuscript text. Authors take full responsibility for the accuracy, originality, and scientific content of this work.

\bibliographystyle{IEEEtran}
\bibliography{cmearrivals}

@ARTICLE{Guhathakurta2021JSWSC,
       author = {{Guhathakurta}, Madhulika},
        title = "{Everyday space weather}",
      journal = {Journal of Space Weather and Space Climate},
         year = 2021,
        month = may,
       volume = {11},
          eid = {36},
        pages = {36},
          doi = {10.1051/swsc/2021019},
       adsurl = {https://ui.adsabs.harvard.edu/abs/2021JSWSC..11...36G}
}

@ARTICLE{Brueckner1995SolarPhysics,
       author = {{Brueckner}, G.~E. and {Howard}, R.~A. and {Koomen}, M.~J. and {Korendyke}, C.~M. and {Michels}, D.~J. and {Moses}, J.~D. and {Socker}, D.~G. and {Dere}, K.~P. and {Lamy}, P.~L. and {Llebaria}, A. and {Bout}, M.~V. and {Schwenn}, R. and {Simnett}, G.~M. and {Bedford}, D.~K. and {Eyles}, C.~J.},
        title = "{The Large Angle Spectroscopic Coronagraph (LASCO)}",
      journal = {Solar Physics},
         year = 1995,
        month = dec,
       volume = {162},
       number = {1-2},
        pages = {357-402},
          doi = {10.1007/BF00733434},
       adsurl = {https://ui.adsabs.harvard.edu/abs/1995SoPh..162..357B}
}

@ARTICLE{Kaiser2008STEREO,
       author = {{Kaiser}, M.~L. and {Kucera}, T.~A. and {Davila}, J.~M. and {St. Cyr}, O.~C. and {Guhathakurta}, M. and {Christian}, E.},
        title = "{The STEREO Mission: An Introduction}",
      journal = {Space Science Reviews},
         year = 2008,
        month = apr,
       volume = {136},
       number = {1-4},
        pages = {5-16},
          doi = {10.1007/s11214-007-9277-0},
       adsurl = {https://ui.adsabs.harvard.edu/abs/2008SSRv..136....5K}
}

@ARTICLE{Kay2024CMEarrivals,
       author = {{Kay}, C. and {Palmerio}, E. and {Riley}, P. and {Mays}, M.~L. and {Nieves-Chinchilla}, T. and {Romano}, M. and {Collado-Vega}, Y.~M. and {Wiegand}, C. and {Chulaki}, A.},
        title = "{Updating Measures of CME Arrival Time Errors}",
      journal = {Space Weather},
         year = 2024,
        month = jul,
       volume = {22},
       number = {7},
          eid = {e2024SW003951},
        pages = {e2024SW003951},
          doi = {10.1029/2024SW003951},
       adsurl = {https://ui.adsabs.harvard.edu/abs/2024SpWea..2203951K}
}

@ARTICLE{Odstrcil2003solarwind,
       author = {{Odstrcil}, D.},
        title = "{Modeling 3-D solar wind structure}",
      journal = {Advances in Space Research},
         year = 2003,
        month = aug,
       volume = {32},
       number = {4},
        pages = {497-506},
          doi = {10.1016/S0273-1177(03)00332-6},
       adsurl = {https://ui.adsabs.harvard.edu/abs/2003AdSpR..32..497O}
}

@ARTICLE{Pomoell2018EUHFORIA,
       author = {{Pomoell}, Jens and {Poedts}, S.},
        title = "{EUHFORIA: European heliospheric forecasting information asset}",
      journal = {Journal of Space Weather and Space Climate},
         year = 2018,
        month = jun,
       volume = {8},
          eid = {A35},
        pages = {A35},
          doi = {10.1051/swsc/2018020},
       adsurl = {https://ui.adsabs.harvard.edu/abs/2018JSWSC...8A..35P}
}

@ARTICLE{Provornikova2024GAMERAGL,
       author = {{Provornikova}, Elena and {Merkin}, Viacheslav G. and {Vourlidas}, Angelos and {Malanushenko}, Anna and {Gibson}, Sarah E. and {Winter}, Eric and {Arge}, Charles N.},
        title = "{MHD Modeling of a Geoeffective Interplanetary Coronal Mass Ejection with the Magnetic Topology Informed by In Situ Observations}",
      journal = {The Astrophysical Journal},
         year = 2024,
        month = dec,
       volume = {977},
       number = {1},
          eid = {106},
        pages = {106},
          doi = {10.3847/1538-4357/ad83b1},
archivePrefix = {arXiv},
       eprint = {2405.13069},
 primaryClass = {astro-ph.SR},
       adsurl = {https://ui.adsabs.harvard.edu/abs/2024ApJ...977..106P}
}

@ARTICLE{Rudisser2026SpWea..2404537R,
       author = {{R{\"u}disser}, Hannah T. and {Nguyen}, Gautier and {Le Lou{\"e}dec}, Justin and {Davies}, Emma E. and {M{\"o}stl}, Christian},
        title = "{ARCANE - Early Detection of Interplanetary Coronal Mass Ejections}",
      journal = {Space Weather},
         year = 2026,
        month = feb,
       volume = {24},
       number = {2},
          eid = {e2025SW004537},
        pages = {e2025SW004537},
          doi = {10.1029/2025SW004537},
archivePrefix = {arXiv},
       eprint = {2505.09365},
 primaryClass = {physics.space-ph},
       adsurl = {https://ui.adsabs.harvard.edu/abs/2026SpWea..2404537R}
}

@ARTICLE{Stone1998SSRvACE,
       author = {{Stone}, E.~C. and {Frandsen}, A.~M. and {Mewaldt}, R.~A. and {Christian}, E.~R. and {Margolies}, D. and {Ormes}, J.~F. and {Snow}, F.},
        title = "{The Advanced Composition Explorer}",
      journal = {Space Science Reviews},
         year = 1998,
        month = jul,
       volume = {86},
        pages = {1-22},
          doi = {10.1023/A:1005082526237},
       adsurl = {https://ui.adsabs.harvard.edu/abs/1998SSRv...86....1S}
}

@ARTICLE{Schrijver2015AdSpR..55.2745S,
       author = {{Schrijver}, Carolus J. and {Kauristie}, Kirsti and {Aylward}, Alan D. and {Denardini}, Clezio M. and {Gibson}, Sarah E. and {Glover}, Alexi and {Gopalswamy}, Nat and {Grande}, Manuel and {Hapgood}, Mike and {Heynderickx}, Daniel and {Jakowski}, Norbert and {Kalegaev}, Vladimir V. and {Lapenta}, Giovanni and {Linker}, Jon A. and {Liu}, Siqing and {Mandrini}, Cristina H. and {Mann}, Ian R. and {Nagatsuma}, Tsutomu and {Nandy}, Dibyendu and {Obara}, Takahiro and {Paul O'Brien}, T. and {Onsager}, Terrance and {Opgenoorth}, Hermann J. and {Terkildsen}, Michael and {Valladares}, Cesar E. and {Vilmer}, Nicole},
        title = "{Understanding space weather to shield society: A global road map for 2015-2025 commissioned by COSPAR and ILWS}",
      journal = {Advances in Space Research},
         year = 2015,
        month = jun,
       volume = {55},
       number = {12},
        pages = {2745-2807},
          doi = {10.1016/j.asr.2015.03.023},
archivePrefix = {arXiv},
       eprint = {1503.06135},
 primaryClass = {physics.space-ph},
       adsurl = {https://ui.adsabs.harvard.edu/abs/2015AdSpR..55.2745S}
}

@ARTICLE{Scherrer2012SDOHMI,
       author = {{Scherrer}, P.~H. and {Schou}, J. and {Bush}, R.~I. and {Kosovichev}, A.~G. and {Bogart}, R.~S. and {Hoeksema}, J.~T. and {Liu}, Y. and {Duvall}, T.~L. and {Zhao}, J. and {Title}, A.~M. and {Schrijver}, C.~J. and {Tarbell}, T.~D. and {Tomczyk}, S.},
        title = "{The Helioseismic and Magnetic Imager (HMI) Investigation for the Solar Dynamics Observatory (SDO)}",
      journal = {Solar Physics},
         year = 2012,
        month = jan,
       volume = {275},
       number = {1-2},
        pages = {207-227},
          doi = {10.1007/s11207-011-9834-2},
       adsurl = {https://ui.adsabs.harvard.edu/abs/2012SoPh..275..207S}
}

@ARTICLE{Arge:2000:WSA,
       author = {{Arge}, C.~N. and {Pizzo}, V.~J.},
        title = "{Improvement in the prediction of solar wind conditions using near-real time solar magnetic field updates}",
      journal = {Journal of Geophysical Research},
         year = 2000,
        month = may,
       volume = {105},
       number = {A5},
        pages = {10465-10480},
          doi = {10.1029/1999JA000262},
          url = {https://doi.org/10.1029/1999JA000262},
       adsurl = {https://ui.adsabs.harvard.edu/abs/2000JGR...10510465A}
}

@ARTICLE{Arge:2004:WSA,
       author = {{Arge}, C.~N. and {Luhmann}, J.~G. and {Odstrcil}, D. and {Schrijver}, C.~J. and {Li}, Y.},
        title = "{Stream structure and coronal sources of the solar wind during the May 12th, 1997 CME}",
      journal = {Journal of Atmospheric and Solar-Terrestrial Physics},
         year = 2004,
        month = oct,
       volume = {66},
       number = {15-16},
        pages = {1295-1309},
          doi = {10.1016/j.jastp.2004.03.018},
          url = {https://doi.org/10.1016/j.jastp.2004.03.018},
       adsurl = {https://ui.adsabs.harvard.edu/abs/2004JASTP..66.1295A}
}

@ARTICLE{Zhang:2019:GAMERA,
       author = {{Zhang}, Binzheng and {Sorathia}, Kareem A. and {Lyon}, John G. and
         {Merkin}, Viacheslav G. and {Garretson}, Jeffrey S. and
         {Wiltberger}, Michael},
        title = "{GAMERA: A Three-dimensional Finite-volume MHD Solver for Non-orthogonal Curvilinear Geometries}",
      journal = {The Astrophysical Journal Supplement Series},
         year = "2019",
        month = "Sep",
       volume = {244},
       number = {1},
          eid = {20},
        pages = {20},
          doi = {10.3847/1538-4365/ab3a4c},
archivePrefix = {arXiv},
       eprint = {1810.10861},
 primaryClass = {physics.comp-ph},
       adsurl = {https://ui.adsabs.harvard.edu/abs/2019ApJS..244...20Z}
}

@ARTICLE{Merkin:2016:GAMERA_a,
       author = {{Merkin}, V.~G. and {Lyon}, J.~G. and {Lario}, D. and {Arge}, C.~N. and {Henney}, C.~J.},
        title = "{Time-dependent magnetohydrodynamic simulations of the inner heliosphere}",
      journal = {Journal of Geophysical Research (Space Physics)},
         year = 2016,
        month = apr,
       volume = {121},
       number = {4},
        pages = {2866-2890},
          doi = {10.1002/2015JA022200},
          url = {https://doi.org/10.1002/2015JA022200},
       adsurl = {https://ui.adsabs.harvard.edu/abs/2016JGRA..121.2866M}
}

@ARTICLE{Merkin:2016:GAMERA_b,
       author = {{Merkin}, V.~G. and {Lionello}, R. and {Lyon}, J.~G. and {Linker}, J. and {T{\"o}r{\"o}k}, T. and {Downs}, C.},
        title = "{Coupling of Coronal and Heliospheric Magnetohydrodynamic Models: Solution Comparisons and Verification}",
      journal = {The Astrophysical Journal},
         year = 2016,
        month = nov,
       volume = {831},
       number = {1},
          eid = {23},
        pages = {23},
          doi = {10.3847/0004-637X/831/1/23},
          url = {https://doi.org/10.3847/0004-637X/831/1/23},
       adsurl = {https://ui.adsabs.harvard.edu/abs/2016ApJ...831...23M}
}

@ARTICLE{Gibson:1998:GL,
       author = {{Gibson}, S.~E. and {Low}, B.~C.},
        title = "{A Time-Dependent Three-Dimensional Magnetohydrodynamic Model of the Coronal Mass Ejection}",
      journal = {The Astrophysical Journal},
         year = "1998",
        month = "Jan",
       volume = {493},
       number = {1},
        pages = {460-473},
          doi = {10.1086/305107},
       adsurl = {https://ui.adsabs.harvard.edu/abs/1998ApJ...493..460G}
}

@INPROCEEDINGS{Sadykov2026COSPAR,
       author = {{Sadykov}, Viacheslav and {Provornikova}, Elena and {Kempton}, Dustin J. and {Sawadogo}, Djalil and {Paouris}, Evangelos and {Arge}, Charles and {Saba}, Tamima and {Angryk}, Rafal A. and {Martens}, Petrus},
        title = "{Intertwining Physics-Based CME Modeling and Machine Learning for L1 Prediction of Interplanetary Magnetic Field}",
    booktitle = {46th COSPAR Scientific Assembly. Held 1-9 August},
         year = 2026,
       volume = {46},
        month = aug,
        pages = {1141},
       adsurl = {https://ui.adsabs.harvard.edu/abs/2026cosp...46.1141S}
}

@ARTICLE{Sadykov:2026:MLdataset,
       author = {{Sadykov}, Viacheslav and {Provornikova}, Elena and {Kempton}, Dustin J. and {Sawadogo}, Djalil and {Mallick}, Debosmita and {Paouris}, Evangelos and {Arge}, Charles and {Saba}, Tamima and {Angryk}, Rafal A. and {Martens}, Petrus},
        title = "{Science-Ready Dataset of Magnetohydrodynamic Simulations of Earth-Impacting Coronal Mass Ejections}",
      journal = {The Astrophysical Journal Supplement Series},
         year = "2026",
         note = {in preparation}
}

@ARTICLE{Singh2023CMEarrivals,
       author = {{Singh}, Talwinder and {Benson}, Bernard and {Raza}, Syed A.~Z. and {Kim}, Tae K. and {Pogorelov}, Nikolai V. and {Smith}, William P. and {Arge}, Charles N.},
        title = "{Improving the Arrival Time Estimates of Coronal Mass Ejections by Using Magnetohydrodynamic Ensemble Modeling, Heliospheric Imager Data, and Machine Learning}",
      journal = {The Astrophysical Journal},
         year = 2023,
        month = may,
       volume = {948},
       number = {2},
          eid = {78},
        pages = {78},
          doi = {10.3847/1538-4357/acc10a},
archivePrefix = {arXiv},
       eprint = {2302.05588},
 primaryClass = {physics.space-ph},
       adsurl = {https://ui.adsabs.harvard.edu/abs/2023ApJ...948...78S}
}

@ARTICLE{Malanushenko:2020:CoNNs,
       author = {{Malanushenko}, Anna and {Flyer}, Natasha and {Gibson}, Sarah},
        title = "{Convolutional Neural Networks for Predicting the strength of the Near-Earth Magnetic Field Caused by Interplanetary Coronal Mass Ejections}",
      journal = {Frontiers in Astronomy and Space Sciences},
         year = 2020,
        month = sep,
       volume = {7},
          eid = {62},
        pages = {62},
          doi = {10.3389/fspas.2020.00062},
       adsurl = {https://ui.adsabs.harvard.edu/abs/2020FrASS...7...62M}
}
\end{document}